%% file: paper_v3.tex
\documentclass[submission, Phys]{SciPost}
\usepackage{color,amsmath,amssymb,multirow,hyperref,booktabs,graphicx,mathtools} 
\input{declare}

\begin{document}

\begin{center}{\Large \textbf{
Deep-learned Top Tagging with a Lorentz Layer
}}\end{center}

\begin{center}
Anja Butter\textsuperscript{1},
Gregor Kasieczka\textsuperscript{2},
Tilman Plehn\textsuperscript{1}, and
Michael Russell\textsuperscript{1,3}
\end{center}

\begin{center}
{\bf 1} Institut f\"ur Theoretische Physik, Universit\"at Heidelberg, Germany\\
{\bf 2} Institute for Particle Physics, ETH Z\"urich, Switzerland \\
{\bf 3} School of Physics and Astronomy, University of Glasgow, Scotland \\
plehn@uni-heidelberg.de
\end{center}

\begin{center}
\today
\end{center}


\section*{Abstract}
{\bf 
  We introduce a new and highly efficient tagger for hadronically
  decaying top quarks, based on a deep neural network working with
  Lorentz vectors and the Minkowski metric. With its novel machine
  learning setup and architecture it allows us to identify
  boosted top quarks not only from calorimeter towers, but also
  including tracking information. We show how the performance of our 
  tagger compares with QCD-inspired and
  image-recognition approaches and find that it significantly
  increases the performance for strongly boosted top quarks.
}

\vspace{10pt}
\noindent\rule{\textwidth}{1pt}
\tableofcontents\thispagestyle{fancy}
\noindent\rule{\textwidth}{1pt}
\vspace{10pt}

\newpage
\section{Introduction}
\label{sec:intro}

The classification of hadronic objects has become the main driving
force behind machine learning techniques in LHC physics. The task
is to identify the partonic nature of large-area jets or fat
jets. Such jets occur for instance in boosted hadronic decays of Higgs
bosons~\cite{bdrs}, weak gauge bosons~\cite{w_tag}, or top
quarks~\cite{early_top,hopkins,template,heptop1,heptop2,heptop3,heptop4,shower_deco,tagging_review}.

A widely debated, central question is how we can analyze these jet
substructure patterns using a range of machine learning techniques.  An early
example were wavelets, describing patterns of hadronic weak boson
decays~\cite{wavelet_tim,wavelet_monk}. The most frequently used
approach is image recognition applied to calorimeter entries in the
azimuthal angle vs rapidity plane, so-called jet images.  They can be
used to search for hadronic decays of weak
bosons~\cite{slac1,slac2,irvine_w,aussies,gan} or top
quarks~\cite{ann_top,deeptop}, or to distinguish quark-like from
gluon-like jets~\cite{quark_gluon}. Another approach is inspired by
natural language recognition, applied to decays of weak
bosons~\cite{kyle}.\smallskip

Top taggers inspired by image recognition rely on convolutional
networks (CNN)~\cite{deeptop,convnet}, which work well 
for numbers of pixels small enough to be
analyzed by the network.  We have shown that they 
can outperform multi-variate QCD-based taggers, but also that
the CNN learns all the appropriate sub-jet patterns~\cite{deeptop}.  A
major problem arises when we include tracking information with its
much better experimental resolution, leading to too many, too sparsely
distributed active pixels~\cite{quark_gluon}.\smallskip

We propose a new approach to jet substructure using machine learning:
rather than relying on analogies to image or natural language
recognition we analyze the constituents of the fat jet directly, only
using elements of special relativity, namely the Lorentz group and
Minkowski metric, to distinguish signal from background.  For our
\textsc{DeepTopLoLa} tagger we introduce a Combination layer (CoLa)
together with a Lorentz layer (LoLa) and two fully connected layers
forming a novel deep neural network (DNN) architecture. In the
standard setup the input 4-momenta correspond to calorimeter
towers~\cite{canadians}. Our \textsc{DeepTopLoLa} tagger can be
extended to include tracking information with its much finer
resolution than the calorimeter granularity.  For any image-based
convolutional network the significantly different resolution of
calorimeter and tracker poses a serious problem.

This flexible setup allows us to study how much performance gain
tracking information actually gives. Moreover, it means that
\textsc{DeepTopLoLa} can be immediately included in state-of-the art
ATLAS and CMS analyses and can be combined with $b$-tagging.\smallskip

In this letter we first introduce our new machine learning
setup. Using standard fat jets from hadronic top decays we compare its
performance to multivariate QCD-inspired tagging and an image-based
convolutional network~\cite{deeptop}. We then extend the tagger to
include particle flow information~\cite{particleflow} and estimate the performance gain
compared to calorimeter information for mildly boosted and strongly
boosted top quarks.

\section{Tagger} 

\begin{figure}[t]
\begin{center}
\includegraphics[width=0.45\textwidth]{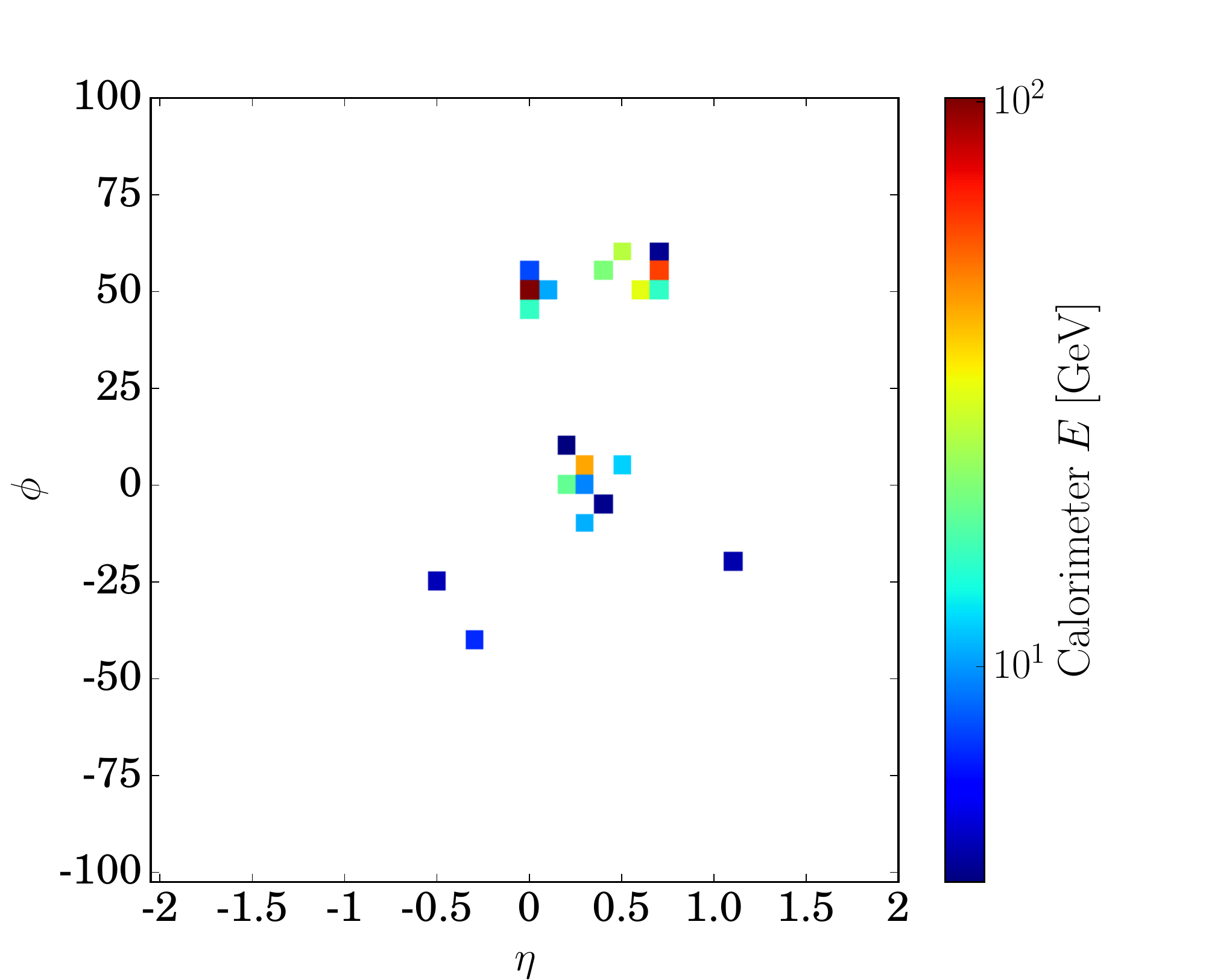}
\end{center}
\vspace*{-6mm}
\caption{Jet image (generated with \textsc{Pythia} and
  \textsc{Delphes}) illustrating a signal event, showing 20 of the jet
  constituent 4-vectors $k_{\mu,i}$. The color scale shows the energy
  of the constituents with a minimum energy threshold $k_0 > 1$~GeV.}
\label{fig:image}
\end{figure}

The basic constituents entering any subjet analysis are a set of $N$
measured 4-vectors sorted by $p_{\textrm{T}}$, for example organized
as the matrix
\begin{align}
( k_{\mu,i} ) = 
\begin{pmatrix}
k_{0,1} &  k_{0,2} & \cdots &  k_{0,N} \\
k_{1,1} &  k_{1,2} & \cdots &  k_{1,N} \\ 
k_{2,1} &  k_{2,2} & \cdots &  k_{2,N} \\ 
k_{3,1} &  k_{3,2} & \cdots &  k_{3,N}  
\end{pmatrix} \; .
\label{eq:def_input}
\end{align}
We show a typical jet image for a hadronic top decay in
Fig.~\ref{fig:image}, indicating that the calorimeters entries of a
typical top decay form a sparsely filled image. A standard approach to
this problem in machine learning are graph convolutional
networks~\cite{graph_cnn}, where such sparse sets of objects are
evaluated as nodes with a learnable distance metric. We further
develop this approach based on the known space-time symmetry structure
linking 4-vectors.

\subsection{Combination layer} 

Our tagger consists of two physics-inspired modules. As a first step,
we multiply the 4-vectors from Eq.\eqref{eq:def_input} with a matrix
$C_{ij}$.  Inspired by the treatment of jet clustering in the
non-deterministic Qjets approach~\cite{qjets} this defines our
Combination layer
\begin{align}
k_{\mu,i} \stackrel{\text{CoLa}}{\longrightarrow}
\widetilde{k}_{\mu,j} 
= k_{\mu,i} \; C_{ij} \; .
\end{align}
It returns $M$ combined 4-vectors $\tilde{k}_j$ made out of the $N$
original input 4-vectors, so $i=1~...~N$ and $j=1~...~M$. From many
top tagging tests we known that an efficient tagger needs to find the
mass drops associated with the top decay and the $W$
decay~\cite{heptop1,heptop2,deeptop}. For illustration purposes, we
look at the two corresponding on-shell conditions in our framework,
\begin{alignat}{11}
\tilde{k}_{\mu,1}^2
&= (k_{\mu,1} + k_{\mu,2} + k_{\mu,3})^2 = m_t^2
\notag \\
\tilde{k}_{\mu,2}^2 
&= (k_{\mu,1} + k_{\mu,2} )^2 = m_W^2 \; .
\end{alignat}
They correspond to non-zero entries
\begin{alignat}{11}
C_{11} = C_{21} = C_{31} 
\quad \text{and} \quad 
C_{12} = C_{22} \; .
\label{eq:illustration}
\end{alignat}
In general, the CoLa matrix in our neural network has the trainable
form
\begin{align}
C = 
 \begin{pmatrix}
  1 & 0 & \cdots & 0      & C_{1,N+2} & \cdots & C_{1,M} \\[-2mm]
  0 & 1 &  & \vdots & C_{2,N+2} & \cdots & C_{2,M} \\[-2mm]
  \vdots & \vdots & \ddots & 0 & \vdots &  & \vdots \\
  0 & 0 & \cdots & 1 &C_{N,N+2} & \cdots & C_{N,M} 
 \end{pmatrix} \; .
\label{eq:weights_cola}
\end{align}
It guarantees that the set of $M$ 4-momenta $\tilde{k}_j$ includes
\begin{enumerate}
\itemsep0em 
\item each original momentum $k_i$;
\item a trainable set of $M-N$ linear combinations.
\end{enumerate}
These $\tilde{k}_j$ will be analyzed by a DNN. 

While one could use advanced pre-processing beyond some kind of
ordering of the input 4-momenta, our earlier study~\cite{deeptop}
suggests that this is not necessary. For our numerical study we vary
$N$, the maximum number of jet constituents kept, sorted by
$p_T$. After testing different values for calorimeter cells or
particle-flow objects for moderately or highly boosted tops, we use 15
trainable combinations, or $M = 15+N$ and have checked that
changing $M$ has no effect.

\subsection{Lorentz layer} 

From fundamental theory we know that the
relevant distance measure between two substructure objects is the
Minkowski metric. We use it to construct a weight function which makes
it easier for the DNN to learn the underlying features\footnote{We
  are grateful to Johann Brehmer for pointing out that this approach
  limits us to fat jets far from black holes.}. Since each constituent momentum
is specified uniquely by four degrees of freedom, we can choose a transformation which maps
the constituent 4-vectors to quantities more directly related to physical observables.  To do this, 
we define a Lorentz layer as
the second part of the DNN which first transforms the $M$ 4-vectors
$\tilde{k}_j$ into the same number of measurement-motivated objects $\hat{k}_j$,
\begin{align}
\tilde{k}_j 
\stackrel{\text{LoLa}}{\longrightarrow}
\hat{k}_j = 
 \begin{pmatrix*}[r]
  m^2(\tilde k_j)\\ 
  p_T(\tilde k_j)\\ 
  w^{(E)}_{jm} \,E(\tilde k_m)\\ 
  w^{(d)}_{jm} \, d^2_{jm}\\ 
 \end{pmatrix*} \; ,
\label{eq:lola}
\end{align}
where $d^2_{jm}$ is the Minkowski distance between two four-momenta
$\tilde k_j$ and $\tilde k_m$,
\begin{align}
d^2_{jm} &= (\tilde k_j - \tilde k_m)_{\mu} \; g^{\mu \nu} \; (\tilde k_j - \tilde k_m)_{\nu} \; ,
\label{eq:minkowski}
\end{align}
combined with the matrices of weights $w_{jm}$ updated during the
training of the network. The four entries illustrate different
structures we can include in this Lorentz layer.  The first two
$\hat{k}_j$ map individual $\tilde{k}_j$ onto their invariant mass and
transverse momentum. The invariant mass entry corresponds to the
illustration in Eq.\eqref{eq:illustration}. The third entry constructs
a linear combination of all energies, with a trainable vector of
weights $w^{(E)}_{jm}$ with $m=1~...~M$. Different values of $j$ give
us a set of $M$ copies of this linear combination. The fourth entry
combines all $\tilde{k}_m$ with a fixed $\tilde{k}_j$, including a
trainable vector of weights $w^{(d)}_{jm}$.  We can either sum over or
minimize over the internal index $m$, always keeping the external
index $j$ fixed. For the third entry with the trainable weights
$w^{(E)}_{jm}$ we choose the sum over the internal index. For the last
entry with the weights $w^{(d)}_{jm}$ we improve the performance of
the network by including four copies with independently trainable
weights. Two of these copies sum over the internal index and two of
then minimize over it.

We have checked that neither the exact composition of the $\hat{k}_j$
nor the number of entries in Eq.\eqref{eq:lola} have an effect on the
performance of our tagger. What is important is that we combine the
invariant mass with an energy or transverse momentum and include the
trainable weights.  The first and last entries in Eq.\eqref{eq:lola}
explicitly use the Minkowski distance defined in
Eq.\eqref{eq:minkowski}.  The LoLa objects $\hat{k}_j$ are the input
of the DNN. One can think of them as a rotation in the observable
space, making the relevant information more accessible to the neural
network, so the LoLa should be loss-less, provided the truncation in
the number of input 4-vectors and the selection in Eq.\eqref{eq:lola}
is carefully tested. Finally, the combined set of trainable weights in
Eq.\eqref{eq:weights_cola} and in Eq.\eqref{eq:lola} is large and can
most likely be reduced for a given application. To maintain the
general structure of our approach we decide to not apply this
optimization.

\section{Performance} 

For any proposed new analysis tool, a realistic and convincing
comparison with the state-of-the-art tools is crucial. For our
\textsc{DeepTopLoLa} tagger we compare its performance with a
QCD-inspired top tagger and with an image-based top tagger, both
working on calorimeter entries.

For our comparison we simulate a hadronic $t\bar{t}$ sample and a QCD
di-jet sample with \textsc{Pythia8.2.15}~\cite{pythia} for the 14~TeV
LHC~\cite{samples}. We ignore multi-parton interactions and in
particular pile-up, leaving this aspect to a dedicated study. Several
common approaches of dealing with pile-up~\cite{puppi,pileup} can be
easily combined with our work. For example, the \textsc{DeepTopLoLa}
algorithm can be applied to jet constituents reconstructed using the
\textsc{Puppi} algorithm~\cite{puppi}, where the \textsc{Puppi} weight
for each constituent can be included as an additional parameter in the
training. Alternatively \textsc{DeepTopLoLa} can be used on
constituents of jets after grooming to remove pile-up has been
applied~\cite{pileup}.  Moreover, we assume that our top tagger can be
trained on a pure sample of lepton-hadron top pair events with an
identified leptonic top decay.

All events are passed through the fast detector simulation
\textsc{Delphes3.3.2}~\cite{delphes}, with calorimeter towers of size
$\Delta\eta \times \Delta\phi = 0.1 \times 5^\circ$ and an energy threshold of
1~GeV.  We cluster these towers with \textsc{FastJet3.1.3}~\cite{fastjet}
to anti-$k_T$~\cite{anti_kt} jets with $R=1.5$. This defines a smooth
outer shape and a jet area of the fat jet. The fat jets have to
fulfill $|\eta_\text{fat}| < 1.0$, to guarantee that they are entirely
in the central part of the detector and to justify our calorimeter
tower size. For signal events, we require that the fat jet can be
associated with a true top quark within $\Delta R < 1.2$.
Unlike in our earlier study we do not re-cluster the anti-$k_T$
jet constituents, because we eventually include tracking information
and do not focus on a comparison with QCD-inspired
taggers~\cite{deeptop}.

\subsection{Calorimeter} 

\begin{figure}[t]
\includegraphics[width=0.45\textwidth]{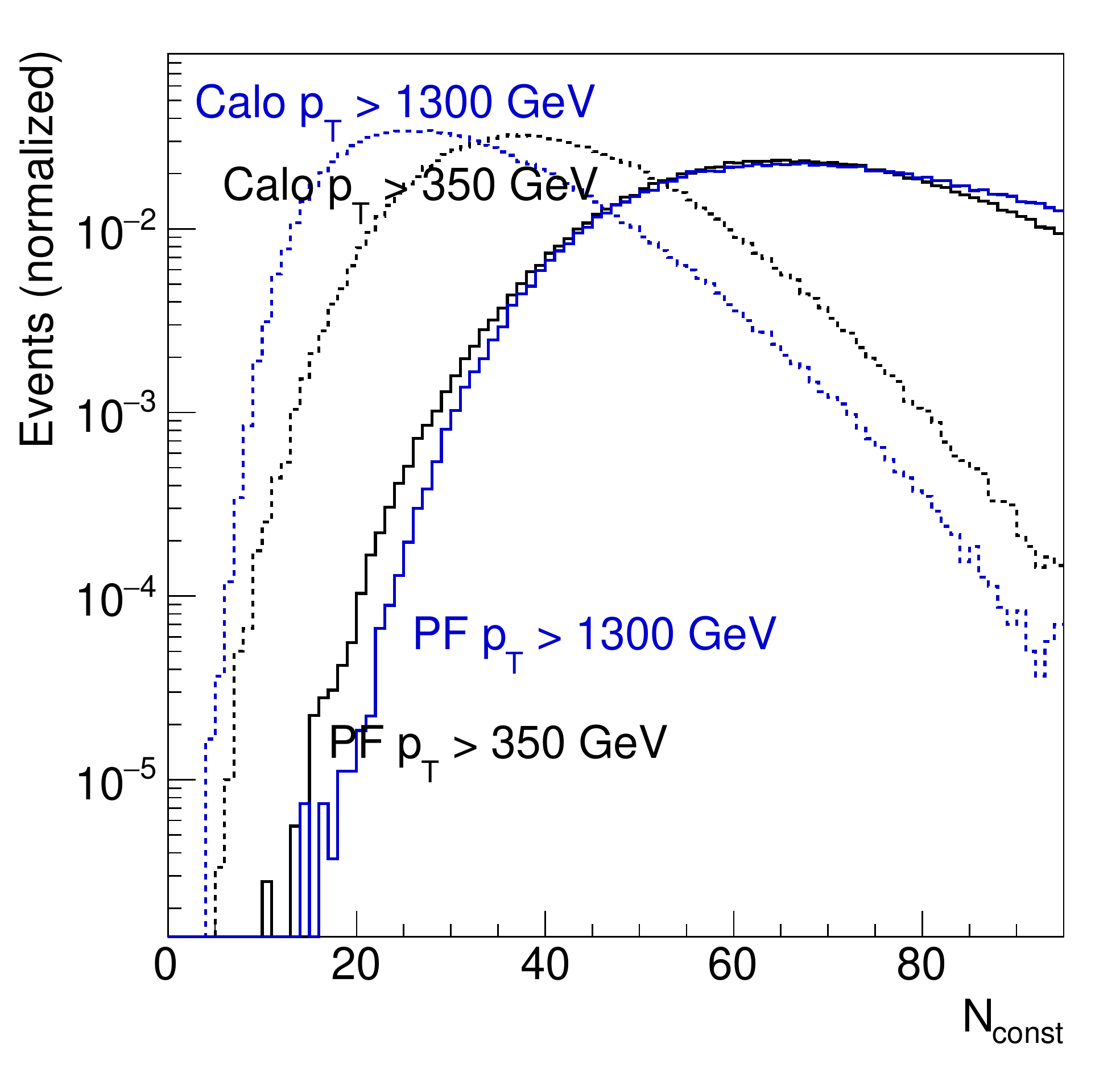}
\hspace*{0.1\textwidth}
\includegraphics[width=0.45\textwidth]{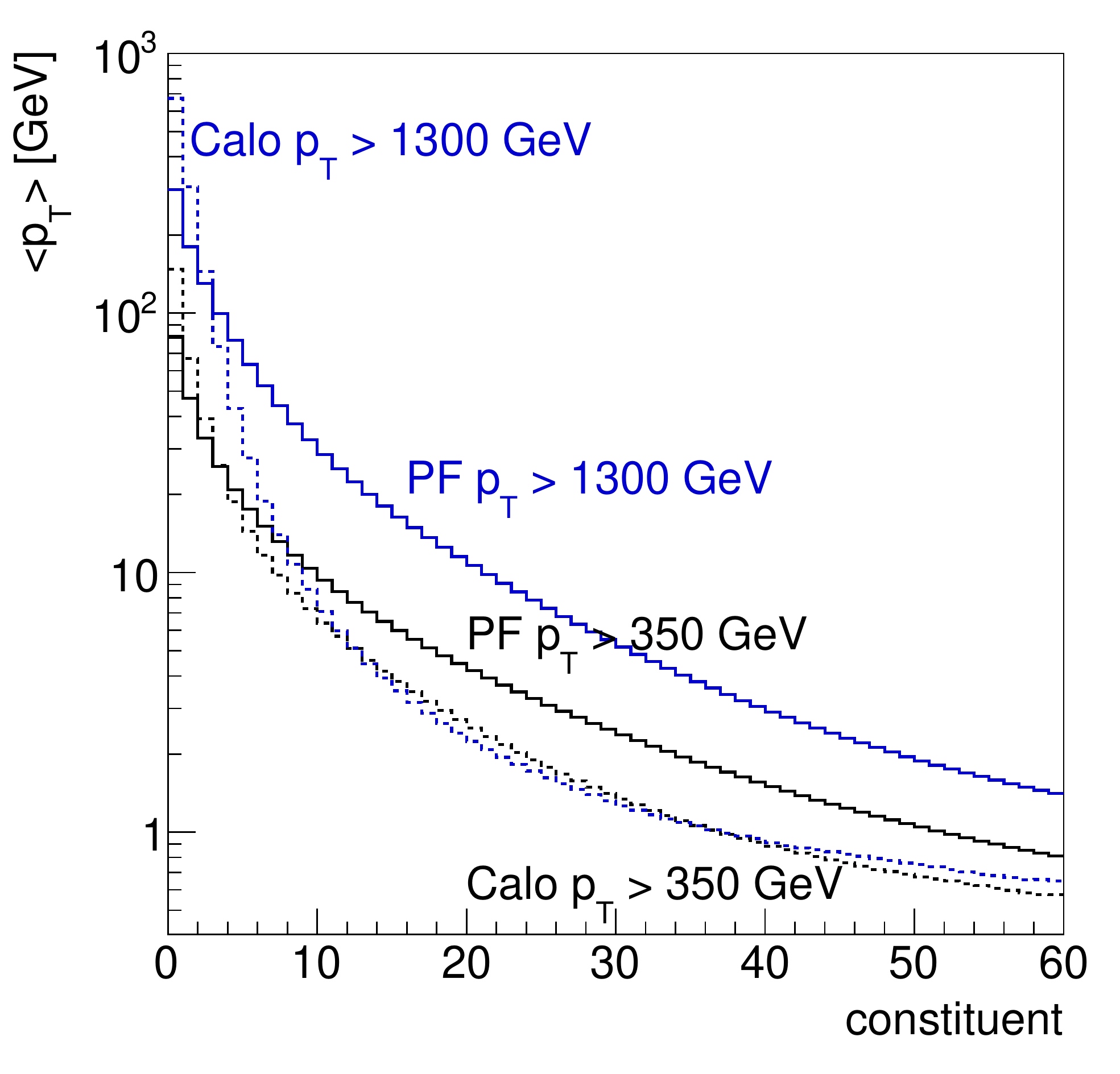}
\vspace*{-6mm}
\caption{Number of constituents $N_{\textrm{const}}$ (left) and mean of the transverse
  momentum (right) of the ranked constituents available as 4-vectors in
  Eq.\eqref{eq:def_input}. We show 4-vectors for the top signal from
  calorimeter cells or jet images (dashed) and from calorimeter and
  tracker information combined through particle flow (solid).}
\label{fig:nconst}
\end{figure}

We consider the two standard ranges, moderately boosted tops available
in Standard Model processes and highly boosted tops in resonance
searches,
\begin{align}
p_{T,\text{fat}} &= 350~...~450~\gev \notag \\
p_{T,\text{fat}} &= 1300~...~1400~\gev \; .
\label{eq:pt_range}
\end{align}
In the left panel of Fig.~\ref{fig:nconst} we show the number of
available calorimeter-based 4-vectors $k_{\mu,i}$, implying that
$N_{\textrm{const}}$ is the maximum number of constituents $N$ we
include in our analysis. In the right panel we show the mean
transverse momentum of the $p_T$-ordered 4-vectors counted as
$i_{\textrm{const}} = 1~...~N_{\textrm{const}}$, for the soft and hard
fat jet selections of Eq.\eqref{eq:pt_range}.  For the soft and hard
selections we have tested values $N=10~...~60$ for the number of
constituents entering our analysis. We find that using the highest
$p_T$ $N=40$ calorimeter constituents completely saturates the tagging
performance. The remaining entries will typically be much softer than
the top decay products and hence carry little signal or background
information from the hard process.\smallskip

For the softer fat jets we use 180,000 signal and 180,000 background
events to train the network, 60,000 events each for tests during
training, and 60,000 events each to estimate the performance. For
technical reason the harder fat jets rely on a $10\%$ smaller sample.

The network includes the CoLa, the LoLa, and two fully connected
hidden layers, one with 100 and one with 50 nodes.  It is trained
using \textsc{Keras}~\cite{keras} with the
\textsc{Theano}~\cite{theano} back-end, and the \textsc{Adam} optimizer.
For the learning rate, i.e. the parameter that determines the step size in the 
numerical minimization of the loss function by gradient descent, we use 0.001. 
Training terminates either after 200 epochs, i.e. single updates 
of the network weights using the full training sample, or when the performance 
on the test sample does not improve for five epochs, typically after several tens of epochs.
\footnote{Using this setup, the training for the softer fat jets takes
  less than 15 minutes in total on a Tesla K80 using a p2.xlarge
  computing instance on Amazon Web Services.}  We independently train
five copies of the network with different initial weight seeds, and compare their performances on the
independent validation sample.\smallskip

\begin{figure}[t]
\begin{center}
\includegraphics[width=0.45\textwidth]{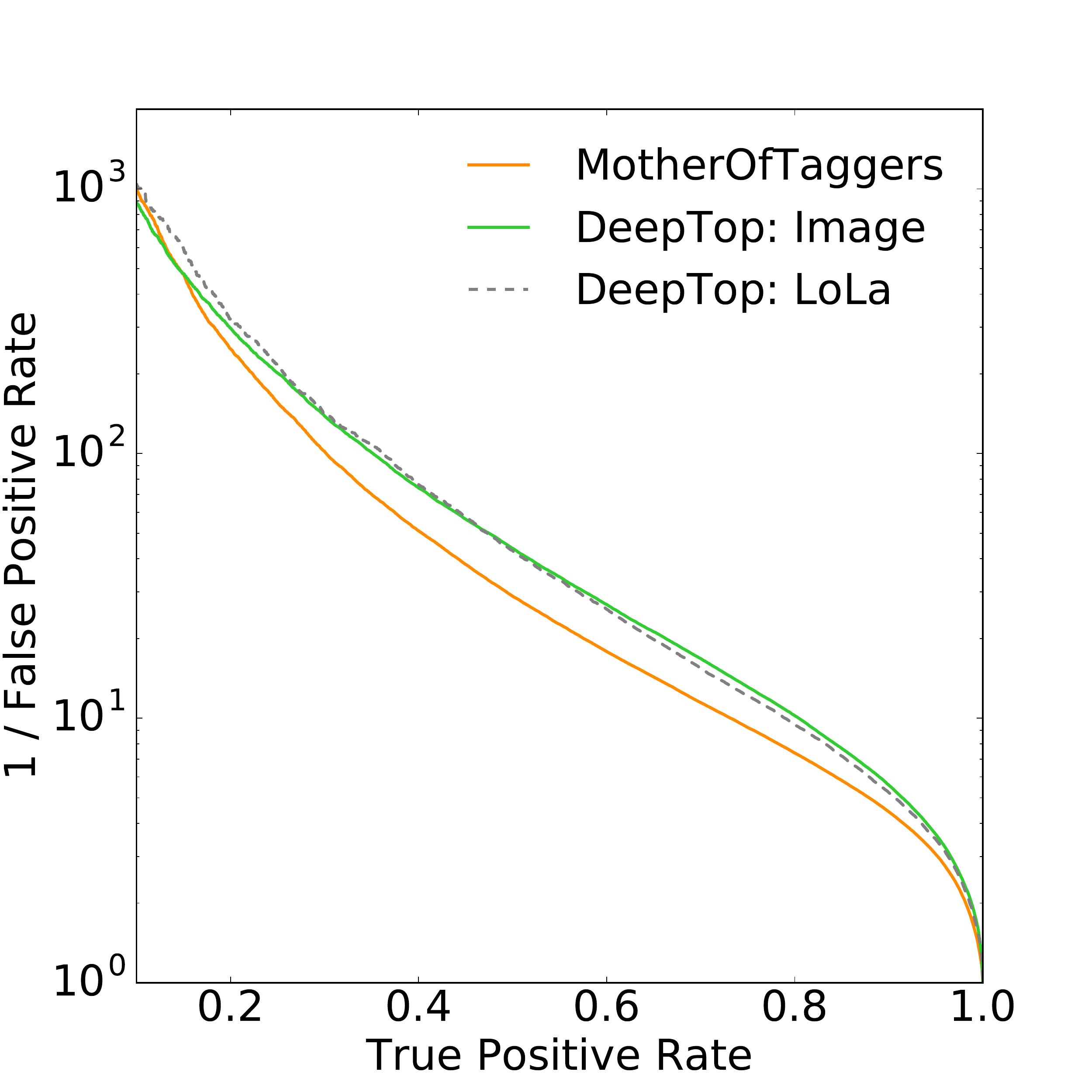}
\end{center}
\vspace*{-6mm}
\caption{ROC curve for the new \textsc{DeepTopLoLa} tagger, compared
  to the QCD-inspired \textsc{MotherOfTaggers} and the image-based
  \textsc{DeepTop} tagger~\cite{deeptop}. In all cases we only use
  calorimeter information for soft fat jets, $p_{T,\text{fat}} =
  350~...~450$~GeV.}
\label{fig:roc_calo}
\end{figure}

Because of a long history of tests and applications on data, top
taggers are especially useful to establish the performance of machine
learning tools.  In Fig.~\ref{fig:roc_calo} we compare our
\textsc{DeepTopLoLa} tagger to earlier benchmarks for the softer of
the two selections in Eq.\eqref{eq:pt_range}: a BDT of a large number
of QCD-inspired observables and the image-based \textsc{DeepTop}
tagger~\cite{deeptop}. The QCD-inspired \textsc{MotherOfTaggers}
consists of a boosted decision tree which includes a large, relatively
well-understood set of observables, which can be linked to a
systematic approach to including sub-jet
correlations~\cite{information}.  It includes the
\textsc{HEPTopTagger} mass drop algorithm~\cite{heptop2} with an
optimal choice of jet size~\cite{heptop4}, different jet masses
including SoftDrop~\cite{softdrop}, as well as
N-subjettiness~\cite{nsubjetti}.  As long as we only include
calorimeter information we cannot expect the new method to
significantly improve over these two approaches. On the other hand,
the number of weights (inputs) of the LoLa-based DNN are lower by a
factor of three to eight (ten to twenty) 
than what is used by the reference convolutional network.
The proposed architecture is 
simpler, more flexible and physics-motivated
but easily matches the convolutional network approach.

\subsection{Learning the Minkowski metric}

A technical challenge related to the Minkowski metric for example in
a graph convolutional network language is that it combines two
different features: two subjets are Minkowski-close if they are
collinear or when one of them is soft ($k_{i,0} \to 0$). Because these
two scenarios correspond to different, but possibly overlapping phase
space regions, they are hard to learn for a DNN.

To see how our \textsc{DeepTopLoLa} tagger deals with this problem and
to test what kind of structures drive the network output, we turn the
problem around and ask the question if the Minkowski metric is really
the feature distinguishing top decays and QCD jets. To this end, we
define the invariant mass $m(\tilde{k}_j)$ and the distance $d_{jm}^2$
in Eq.\eqref{eq:lola} with a trainable diagonal metric.  After
applying a global normalization we find
\begin{align}
g = \text{diag} ( &\quad 0.99 \pm 0.02, 
\\ 
&-1.01 \pm 0.01, -1.01 \pm 0.02, -0.99 \pm 0.02 ) \; , \notag 
\end{align}
where the errors are given by five independently trained copies.  It
is crucial for our physics understanding~\cite{information} that
the distinguishing power of the \textsc{DeepTopLoLa} tagger is indeed
the same mass drop~\cite{bdrs} that drives many QCD-based top
taggers~\cite{heptop1,heptop2} and the image-based top
tagger, as shown in detail in Ref.~\cite{deeptop}.

\subsection{Calorimeter and tracking} 

A standard criticism of the jet image
approach is that the pixelled image removes information from the
original jet. For the calorimeter information alone this is not the
case, because the image pixels are given by the calorimeter
resolution.  However, this identification is not possible for tracking
information, because the tracking resolution of ATLAS and CMS is much
finer than a jet image can realistically
resolve~\cite{quark_gluon}. This makes it hard to in general extend
jet images to particle flow objects and to reliably determine how much
performance can be gained through tracking information.\smallskip

In contrast, for our LoLa-based approach this extension to particle
flow constituents is straightforward: instead of defining one
constituent or 4-vector per calorimeter cell we use all objects
defined by the \textsc{Delphes3} particle flow algorithm in the same
$p_{T,\text{fat}}$ range as in Eq.\eqref{eq:pt_range}. The
fat jet constituents at the particle flow level are different from the
calorimeter case, which implies that for the same $p_{T,\text{fat}}$
range the underlying top quarks are around 5\% softer for fat jets based
on particle flow objects. Nevertheless, defining the signal and
background events using Eq.\eqref{eq:pt_range} still is the best
choice.\smallskip

In Fig.~\ref{fig:nconst} we show the number of constituents for the
calorimeter-level and the particle flow approaches. We see that because 
of the higher precision on the latter, more particle flow objects are resolved on average. 
We also show the mean transverse
momentum for each of these constituents, indicating that the larger
number of particle flow objects at least in part arises from splitting
harder calorimeter entries into several objects at higher resolution.
For our \textsc{DeepTopLoLa} tagger Fig.~\ref{fig:nconst} implies that
we could include more particle flow objects than calorimeter objects
in Eq.\eqref{eq:def_input}. Again, we use $N = 40$ and confirm that an
increase to $N=60$ has no measurable effect on the
performance.\smallskip

Searching for possible improvements to our tagger, we first check that indeed
the top quark kinematics are more precisely measured by the particle
flow objects. However, the observed 5\% improvement, for example in the
resolution of the top transverse momentum, is unlikely to significantly
improve our analysis. 

\begin{figure}[t]
\begin{center}
\includegraphics[width=0.45\textwidth]{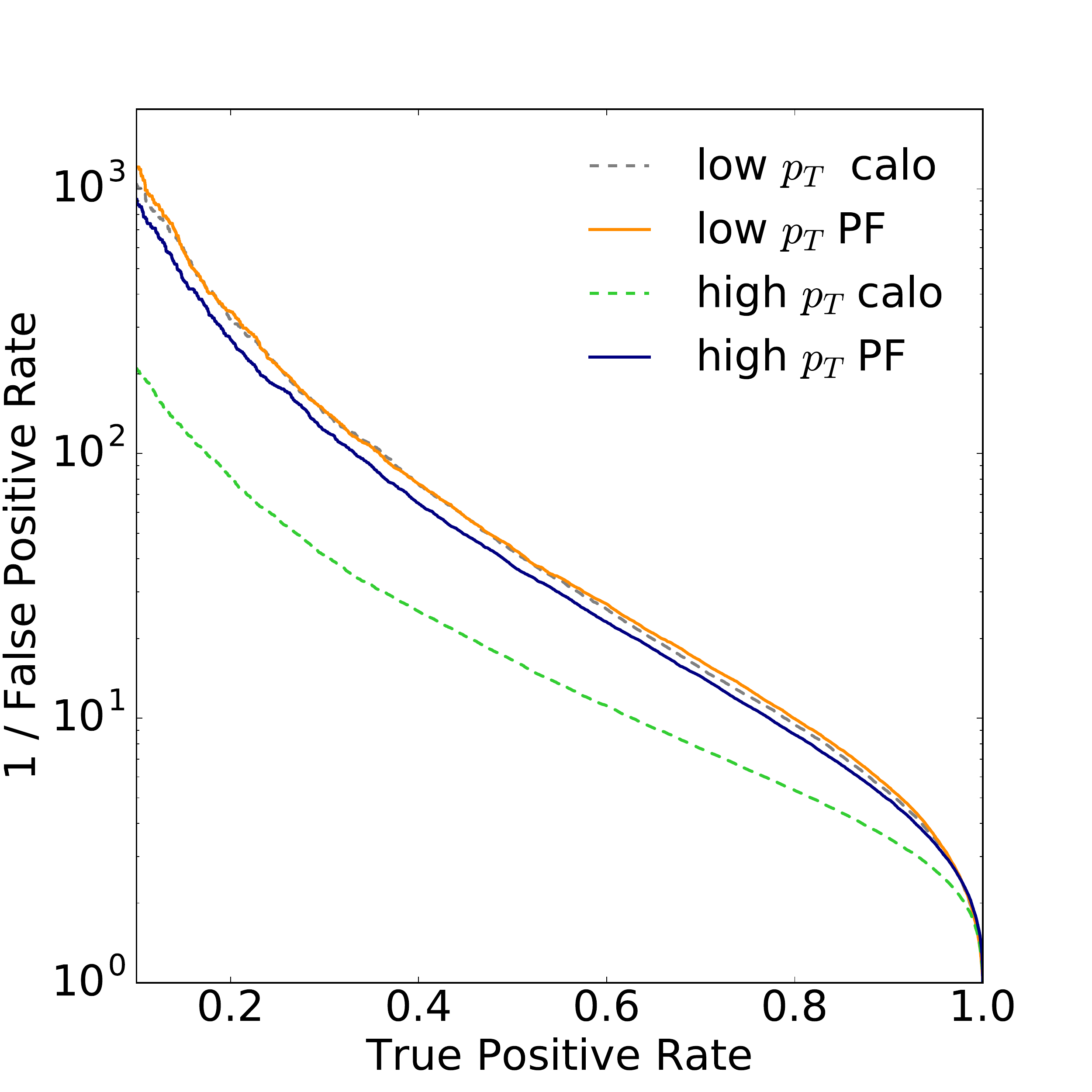}
\end{center}
\vspace*{-6mm}
\caption{ROC curve for the new \textsc{DeepTopLoLa} tagger operating
  on particle flow objects, compared to the its performance operating
  on calorimeter objects.}
\label{fig:roc_flow}
\end{figure}

In Fig.~\ref{fig:roc_flow} we confirm that using the same neural
network for calorimeter and particle flow objects gives hardly any
improvement for moderately boosted tops with $p_{T,\text{fat}} =
350~...~450$~GeV. The situation changes when we train and test our
tagger at larger transverse momenta, $p_{T,\text{fat}} =
1300~...~1400$~GeV.  Here the calorimeter resolution is no longer
sufficient to separate the substructures~\cite{spanno}. For a fixed
signal efficiency the background rejection including particle flow
increases by a factor of two to three.

\section{Conclusions} 

Based on a deep neural network working on
Lorentz vectors of jet constituents we have built the new, simple, and
flexible \textsc{DeepTopLoLa} tagger. It includes a Combination layer
mimicking QCD-inspired jet recombination, a Lorentz layer translating
the 4-vectors into appropriate kinematic observables, and two fully connected
layers. The 4-vector input is not limited to a single detector output but 
allows us to add more information about a subjet object in a 
straightforward manner.

We have compared the tagging performance to QCD-inspired taggers and
to image-based convolutional network taggers using only calorimeter
information for moderately boosted top quarks~\cite{deeptop}.
Figure~\ref{fig:roc_calo} shows that the new tagger is competitive
with either of these alternative approaches. Because we consider it 
crucial to control what machine learning methods actually 
exploit~\cite{information} we not only compared the \textsc{DeepTopLola} 
performance to an established QCD-inspired tagger~\cite{deeptop}, but 
also confirmed that the
Minkowski metric related to a mass drop condition indeed drives the
signal and background distinction.

Finally, we have used our tagger on particle flow objects, combining
calorimeter and tracker information at their respective full
experimental resolution. We have found that while for moderately
boosted top quarks the performance gain from the tracker is negligible, it
makes a big difference for strongly boosted top quarks. 

The coverage of the full transverse momentum range and the possibility
to include $b$-tagging through the tracking information should make
the \textsc{DeepTopLoLa} tagger an excellent starting point to employ
machine learning as the standard in ATLAS and CMS subjets analyses. It
also opens a wide range of applications based on 4-vectors describing
structures like for example matrix elements or phase space.

\bigskip
\begin{center} \textbf{Acknowledgments} \end{center}

First, we would like to thank Anke
Biek\"otter for her help on tracking simulation.  A.B acknowledges
support form the \textsl{Heidelberg Graduate School for Fundamental
  Physics} and the DFG research training group \textsl{Particle
  Physics Beyond the Standard Model}. M.R. was supported by the
European Union Marie Curie Research Training Network MCnetITN, under
contract PITN-GA-2012-315877. Our work was supported by a grant from
the Swiss National Supercomputing Centre (CSCS) under project D61.


\end{document}

%% file: declare.tex
\newcommand{\arxiv}[1]{\href{http://arxiv.org/abs/#1}{arXiv:#1}}
\newcommand{\urlx}[1]{\href{#1}{#1}}

\newcommand\one{\leavevmode\hbox{\small1\normalsize\kern-.33em1}}

\newcommand{\gev}{\text{GeV}}

\def\slashchar#1{\setbox0=\hbox{$#1$}           
   \dimen0=\wd0                                 
   \setbox1=\hbox{/} \dimen1=\wd1               
   \ifdim\dimen0>\dimen1                        
      \rlap{\hbox to \dimen0{\hfil/\hfil}}      
      #1                                        
   \else                                        
      \rlap{\hbox to \dimen1{\hfil$#1$\hfil}}   
      /                                         
   \fi}

\newcommand{\eg}{\textsl{e.g.}\;}

\newcommand{\be}{\begin{eqnarray*}}
\newcommand{\ee}{\end{eqnarray*}}

\newcommand{\bee}{\begin{eqnarray}}
\newcommand{\eee}{\end{eqnarray}}
\newcommand{\beeq}{\begin{equation}}
\newcommand{\eeeq}{\end{equation}}




%% file: paper_v3.bbl
\begin{thebibliography}{99}

\bibitem{bdrs}
  J.~M.~Butterworth, A.~R.~Davison, M.~Rubin and G.~P.~Salam,
  Phys.\ Rev.\ Lett.\  {\bf 100}, 242001 (2008)
  \doi{10.1103/PhysRevLett.100.242001}
  [\arxiv{0802.2470} [hep-ph]].

\bibitem{w_tag}
  M.~H.~Seymour,
  Z.\ Phys.\  C {\bf 62}, 127 (1994);
  \doi{10.1007/BF01559532}
  J.~M.~Butterworth, B.~E.~Cox and J.~R.~Forshaw,
  Phys.\ Rev.\  D {\bf 65}, 096014 (2002)
  \doi{10.1103/PhysRevD.65.096014}
  [\arxiv{hep-ph/0201098}];
  Y.~Cui, Z.~Han and M.~D.~Schwartz,
  Phys.\ Rev.\ D {\bf 83}, 074023 (2011)
  \doi{10.1103/PhysRevD.83.074023}
  [\arxiv{1012.2077} [hep-ph]].

\bibitem{early_top}
 W.~Skiba and D.~Tucker-Smith,
  Phys.\ Rev.\  D {\bf 75}, 115010 (2007)
  \doi{10.1103/PhysRevD.75.115010}
  [\arxiv{hep-ph/0701247}];
 B.~Holdom,
  JHEP {\bf 0703}, 063 (2007)
  \doi{10.1103/PhysRevD.75.115010}
  [\arxiv{hep-ph/0702037}];
 M.~Gerbush, T.~J.~Khoo, D.~J.~Phalen, A.~Pierce and D.~Tucker-Smith,
  Phys.\ Rev.\  D {\bf 77}, 095003 (2008)
  \doi{10.1103/PhysRevD.77.095003}
  [\arxiv{0710.3133} [hep-ph]].

\bibitem{hopkins}
 D.~E.~Kaplan, K.~Rehermann, M.~D.~Schwartz and B.~Tweedie,
  \doi{10.1103/PhysRevLett.101.142001}
  Phys.\ Rev.\ Lett.\  {\bf 101}, 142001 (2008)
  [\arxiv{0806.0848} [hep-ph]].

\bibitem{template}
 L.~G.~Almeida, S.~J.~Lee, G.~Perez, I.~Sung and J.~Virzi,
  Phys.\ Rev.\  D {\bf 79}, 074012 (2009)
  \doi{10.1103/PhysRevD.79.074012}
  [\arxiv{0810.0934} [hep-ph]];
 L.~G.~Almeida, S.~J.~Lee, G.~Perez, G.~F.~Sterman, I.~Sung and J.~Virzi,
  Phys.\ Rev.\ D\ {\bf 79}, 074017  (2009)
  \doi{10.1103/PhysRevD.79.074017}
  [\arxiv{0807.0234} [hep-ph]];
 L.~G.~Almeida, S.~J.~Lee, G.~Perez, G.~Sterman, I.~Sung,
  Phys.\ Rev.\  {\bf D82}, 054034 (2010)
  \doi{10.1103/PhysRevD.82.054034}
  [\arxiv{1006.2035} [hep-ph]];
  M.~Backovic and J.~Juknevich,
  Comput.\ Phys.\ Commun.\  {\bf 185}, 1322 (2014)
  \doi{10.1016/j.cpc.2013.12.018}
  [\arxiv{1212.2978}].

\bibitem{heptop1}
  T.~Plehn, G.~P.~Salam and M.~Spannowsky,
  \doi{10.1103/PhysRevLett.104.111801}
  Phys.\ Rev.\ Lett.\  {\bf 104}, 111801 (2010)
  [\arxiv{0910.5472} [hep-ph]].

\bibitem{heptop2}
 T.~Plehn, M.~Spannowsky, M.~Takeuchi, and D.~Zerwas,
  JHEP {\bf 1010}, 078 (2010)
  \doi{10.1007/JHEP10(2010)078}
  [\arxiv{1006.2833} [hep-ph]].
  \urlx{http://www.thphys.uni-heidelberg.de/~plehn}

\bibitem{heptop3}
 T.~Plehn, M.~Spannowsky and M.~Takeuchi,
  Phys.\ Rev.\ D {\bf 85}, 034029 (2012)
  \doi{10.1103/PhysRevD.85.034029}
  [\arxiv{1111.5034} [hep-ph]];
  C.~Anders, C.~Bernaciak, G.~Kasieczka, T.~Plehn and T.~Schell,
  Phys.\ Rev.\ D {\bf 89}, no. 7, 074047 (2014)
  \doi{10.1103/PhysRevD.89.074047}
  [\arxiv{1312.1504} [hep-ph]].

\bibitem{heptop4} 
  G.~Kasieczka, T.~Plehn, T.~Schell, T.~Strebler and G.~P.~Salam,
  JHEP {\bf 1506}, 203 (2015)
  \doi{10.1007/JHEP06(2015)203}
  [\arxiv{1503.05921} [hep-ph]].

\bibitem{shower_deco}
 D.~E.~Soper and M.~Spannowsky,
  Phys.\ Rev.\ D {\bf 87}, 054012 (2013)
  \doi{10.1103/PhysRevD.87.054012}
  [\arxiv{1211.3140} [hep-ph]].

\bibitem{tagging_review}
for a review see \eg
 A.~Abdesselam  {\it et al.},
  Eur.\ Phys.\ J.\ C {\bf 71}, 1661 (2011)
  \doi{10.1140/epjc/s10052-011-1661-y}
  [\arxiv{1012.5412} [hep-ph]];
 T.~Plehn and M.~Spannowsky,
  J.\ Phys.\ G {\bf 39}, 083001 (2012)
  \doi{10.1088/0954-3899/39/8/083001}
  [\arxiv{1112.4441} [hep-ph]];
 A.~Altheimer {\it et al.},
  Eur.\ Phys.\ J.\ C {\bf 74}, no. 3, 2792 (2014)
  \doi{10.1140/epjc/s10052-014-2792-8}
  [\arxiv{a1311.2708} [hep-ex]];
  S.~Sch\"atzel,
  Eur.\ Phys.\ J.\ C {\bf 75}, no. 9, 415 (2015)
  \doi{10.1140/epjc/s10052-015-3636-x}
  [\arxiv{1403.5176} [hep-ex]].

\bibitem{wavelet_tim}
  V.~Rentala, W.~Shepherd and T.~M.~P.~Tait,
  JHEP {\bf 1408}, 042 (2014)
  \doi{10.1007/JHEP08(2014)042}
  [\arxiv{1404.1929} [hep-ph]].

\bibitem{wavelet_monk}
  J.~W.~Monk,
  [\arxiv{1405.5008} [hep-ph]].

\bibitem{slac1} 
  J.~Cogan, M.~Kagan, E.~Strauss and A.~Schwarztman,
  JHEP {\bf 1502}, 118 (2015)
  \doi{10.1007/JHEP02(2015)118}
  [\arxiv{1407.5675} [hep-ph]].

\bibitem{slac2}
  L.~de Oliveira, M.~Kagan, L.~Mackey, B.~Nachman and A.~Schwartzman,
  JHEP {\bf 1607}, 069 (2016)
  \doi{10.1007/JHEP07(2016)069}
  [\arxiv{1511.05190} [hep-ph]].

\bibitem{irvine_w}
  P.~Baldi, K.~Bauer, C.~Eng, P.~Sadowski and D.~Whiteson,
  Phys.\ Rev.\ D {\bf 93}, no. 9, 094034 (2016)
  \doi{10.1103/PhysRevD.93.094034}
  [\arxiv{1603.09349} [hep-ex]].

\bibitem{aussies}
  J.~Barnard, E.~N.~Dawe, M.~J.~Dolan and N.~Rajcic,
  Phys.\ Rev.\ D {\bf 95}, no. 1, 014018 (2017)
  \doi{10.1103/PhysRevD.95.014018}
  [\arxiv{1609.00607} [hep-ph]].

\bibitem{gan}
  L.~de Oliveira, M.~Paganini and B.~Nachman,
  Comput.\ Softw.\ Big Sci.\  {\bf 1}, no. 1, 4 (2017)
  \doi{10.1007/s41781-017-0004-6}
  [\arxiv{1701.0592} [stat.ML]].

\bibitem{ann_top}
  L.~G.~Almeida, M.~Backovic, M.~Cliche, S.~J.~Lee and M.~Perelstein,
  JHEP {\bf 1507}, 086 (2015)
  \doi{10.1007/JHEP07(2015)086}
  [\arxiv{1501.05968} [hep-ph]].

\bibitem{deeptop}
  G.~Kasieczka, T.~Plehn, M.~Russell and T.~Schell,
  JHEP {\bf 1705}, 006 (2017)
  \doi{10.1007/JHEP05(2017)006}
  [\arxiv{1701.08784} [hep-ph]].

\bibitem{quark_gluon}
  P.~T.~Komiske, E.~M.~Metodiev and M.~D.~Schwartz,
  JHEP {\bf 1701}, 110 (2017)
  \doi{10.1007/JHEP01(2017)110}
  [\arxiv{1612.01551} [hep-ph]].
  
\bibitem{kyle} 
  G.~Louppe, K.~Cho, C.~Becot and K.~Cranmer,
  \arxiv{1702.00748} [hep-ph].

\bibitem{convnet}
  see \eg
  Y.~LeCun, Y.~Bengio, and G.~Hinton, Geoffrey,
  Nature {\bf 7553}, 436 (2015).

\bibitem{canadians} 
  for some similar ideas see \eg
  J.~Pearkes, W.~Fedorko, A.~Lister and C.~Gay,
  \arxiv{1704.02124} [hep-ex].

  \bibitem{particleflow}
  A.~M.~Sirunyan {\it et al.} [CMS Collaboration],
  JINST {\bf 12} (2017) no.10,  P10003
  \doi{10.1088/1748-0221/12/10/P10003}
  [\arxiv{1706.04965} [physics.ins-det]].

\bibitem{qjets}
  S.~D.~Ellis, A.~Hornig, T.~S.~Roy, D.~Krohn and M.~D.~Schwartz,
  Phys.\ Rev.\ Lett.\  {\bf 108}, 182003 (2012)
  \doi{10.1103/PhysRevLett.108.182003}
  [\arxiv{1201.1914} [hep-ph]].

\bibitem{graph_cnn}
  see e.g.
  J.~Bruna, W~Zaremba, A.~Szlam, and Y.~LeCun,
  [\arxiv{1312.6203}];
  M.~Henaff, J.~Bruna, and Y~LeCun,
  [\arxiv{1506.05163}];
  M.~Niepert, M~Ahmed, and K~Kutzkov,
  ICML 2016
  [\arxiv{1605.05273}].
  T.~N.~Kipf,
  [\arxiv{1609.02907}].


\bibitem{pythia}
  T.~Sj\"ostrand, S.~Ask, J.~R.~Christiansen, R.~Corke, N.~Desai, P.~Ilten, S.~Mrenna and S.~Prestel {\it et al.},
  [\arxiv{1410.3012} [hep-ph]].

\bibitem{samples}
 our event samples are available for performance studies upon request.

\bibitem{puppi}
  D.~Bertolini, P.~Harris, M.~Low and N.~Tran,
  JHEP {\bf 1410}, 059 (2014)
  \doi{10.1007/JHEP10(2014)059}
  [\arxiv{1407.6013} [hep-ph]].

\bibitem{pileup}
  M.~Cacciari, G.~P.~Salam and G.~Soyez,
  Eur.\ Phys.\ J.\ C {\bf 75}, no. 2, 59 (2015)
  \doi{10.1140/epjc/s10052-015-3267-2}
  [\arxiv{1407.0408} [hep-ph]];
  P.~T.~Komiske, E.~M.~Metodiev, B.~Nachman and M.~D.~Schwartz,
  JHEP {\bf 1712}, 051 (2017)
  \doi{10.1007/JHEP12(2017)051}
  [\arxiv{1707.08600} [hep-ph]].

\bibitem{delphes} 
  J.~de Favereau {\it et al.} [DELPHES 3 Collaboration],
  JHEP {\bf 1402}, 057 (2014)
  \doi{10.1007/JHEP02(2014)057}
  [\arxiv{1307.6346} [hep-ex]].

\bibitem{fastjet}
 M.~Cacciari and G.~P.~Salam,
  Phys.\ Lett.\  B {\bf 641}, 57 (2006)
  \doi{10.1016/j.physletb.2006.08.037}
  [\arxiv{hep-ph/0512210}];
 M.~Cacciari, G.~P.~Salam and G.~Soyez,
  Eur.\ Phys.\ J.\ C {\bf 72}, 1896 (2012)
  \doi{10.1140/epjc/s10052-012-1896-2}
  [\arxiv{1111.6097} [hep-ph]].
  \urlx{http://fastjet.fr}

\bibitem{anti_kt}
  M.~Cacciari, G.~P.~Salam and G.~Soyez,
  JHEP {\bf 0804}, 063 (2008)
  \doi{10.1088/1126-6708/2008/04/063}
  [\arxiv{0802.1189} [hep-ph]].

\bibitem{keras}
  F.~Chollet,
  \urlx{https://github.com/fchollet/keras} (2015).

\bibitem{theano}
  Theano Development Team,
  [\arxiv{1605.02688} [cs.SC]].

\bibitem{information} 
  K.~Datta and A.~Larkoski,
  JHEP {\bf 1706}, 073 (2017)
  \doi{10.1007/JHEP06(2017)073}
  \arxiv{1704.08249} [hep-ph].

\bibitem{softdrop}
  A.~J.~Larkoski, S.~Marzani, G.~Soyez and J.~Thaler,
  JHEP {\bf 1405}, 146 (2014)
  \doi{10.1007/JHEP05(2014)146}
  [\arxiv{1402.2657} [hep-ph]].

\bibitem{nsubjetti}
  J.~Thaler and K.~Van Tilburg,
  JHEP {\bf 1202}, 093 (2012)
  \doi{10.1007/JHEP02(2012)093}
  [\arxiv{1108.2701} [hep-ph]].

\bibitem{spanno} 
  S.~Schaetzel and M.~Spannowsky,
  Phys.\ Rev.\ D {\bf 89}, no. 1, 014007 (2014)
  \doi{10.1103/PhysRevD.89.014007}
  [\arxiv{1308.0540} [hep-ph]].

\end{thebibliography}
